\documentclass[%
twocolumn,
superscriptaddress,
 amsmath,amssymb,
 aps,prb,
]{revtex4-2}

\usepackage{graphicx}
\usepackage{dcolumn}
\usepackage{bm}
\usepackage{color}
\usepackage{ulem}
\usepackage{xspace}
\usepackage{multirow}
\usepackage[unicode=true,colorlinks=true,linkcolor=blue,citecolor=blue,urlcolor=blue]{hyperref}
\usepackage{physics}
\usepackage{braket}
\usepackage{here}
\usepackage[utf8]{inputenc}

\usepackage{comment}
\usepackage{newtxtext}
\usepackage{newtxmath}
\usepackage{glossaries}
\usepackage{siunitx}
\usepackage[T1]{fontenc}
\setacronymstyle{long-short}
\newacronym{qha}{QHA}{quasiharmonic approximation}
\newacronym{ifc}{IFC}{interatomic force constant}
\newacronym{zsisa}{ZSISA}{zero static internal stress approximation}
\newacronym{v-z}{v-ZSISA}{volumetric ZSISA}
\newacronym{dft}{DFT}{density functional theory}
\newacronym{asr}{ASR}{acoustic sum rule}

\DeclareUnicodeCharacter{2009}{\,} 

\definecolor{green}{rgb}{0,0.6,0.1}

\begin{document}

\preprint{APS/123-QED}

\title{Continuous crossover between insulating ferroelectrics and the polar metals: \textit{Ab initio} calculation of structural phase transitions of Li$B$O$_3$ ($B$ = Ta, W, Re, Os)}
\author{Ryota Masuki}
\email{masuki-ryota774@g.ecc.u-tokyo.ac.jp}
\affiliation{
Department of Applied Physics, The University of Tokyo,7-3-1 Hongo, Bunkyo-ku, Tokyo 113-8656, Japan
}

\author{Takuya Nomoto}
\email{nomoto@ap.t.u-tokyo.ac.jp}
\affiliation{
Research Center for Advanced Science and Technology, The University of Tokyo,
4-6-1 Komaba Meguro-ku, Tokyo 153-8904, Japan
}

\author{Ryotaro Arita}
\email{arita@riken.jp}
\affiliation{
Research Center for Advanced Science and Technology, The University of Tokyo,
4-6-1 Komaba Meguro-ku, Tokyo 153-8904, Japan
}
\affiliation{ 
RIKEN Center for Emergent Matter Science, 2-1 Hirosawa, Wako, Saitama 351-0198, Japan 
}
\author{Terumasa Tadano}
\email{TADANO.Terumasa@nims.go.jp }
\affiliation{ 
CMSM, National Institute for Materials Science (NIMS), 1-2-1 Sengen, Tsukuba, Ibaraki 305-0047, Japan
}


\date{\today}

\begin{abstract}
Inspired by the recent discovery of a new polar metal LiReO$_3$ by K. Murayama \textit{et al}, we calculate the temperature($T$)-dependent crystal structures of Li$B$O3 with $B$ = Ta, W, Re, Os, using the self-consistent phonon (SCPH) theory. We have reproduced the experimentally observed polar-nonpolar structural phase transitions and the transition temperatures ($T_c$) of LiTaO$_3$, LiReO$_3$, and LiOsO$_3$. From the calculation, we predict that LiWO$_3$ is a polar metal, which is yet to be tested experimentally.
Upon doping electrons to the insulating LiTaO$_3$, the predicted $T_c$ is quickly suppressed and approaches those of the polar metals. Thus, there is a continuous crossover between ferroelectric insulators and polar metals if we dope electrons to the polar metals. Investigating the detailed material dependence of the interatomic force constants (IFCs), we explicitly show that the suppression of $T_c$ in polar metals can be ascribed to the screening of the long-range Li-O interaction, which is caused by the presence of the itinerant electrons.
\end{abstract}

\maketitle


\section{Introduction}
\label{Sec_Introduction}
Polar metals, metals that show ferroelectric-like phase transition in terms of crystal symmetry, have attracted considerable interest since LiOsO$_3$ was experimentally identified as a polar metal for the first time~\cite{Shi2013}.
With the coexistence of broken inversion symmetry and metallic conductivity, polar metals can host emergent phenomena~\cite{doi:10.1146/annurev-matsci-080921-105501} such as superconductivity~\cite{Rischau2017, PhysRevB.95.100501}, enhanced thermoelectricity~\cite{doi:10.1126/sciadv.1601378, PhysRevB.100.195130}, nontrivial topology~\cite{PhysRevMaterials.2.051201, Li2017}, and multiferroicity~\cite{PhysRevB.96.235415}. 

However, only a few materials have been discovered~\cite{doi:10.1146/annurev-matsci-080921-105501, Zhou_2020, PhysRevMaterials.7.010301}, although half a century has passed since its first proposal~\cite{PhysRevLett.14.217}. This is because the long-range Coulomb interaction between local dipoles, which causes the ferroelectric instabilities in insulators~\cite{doi:10.1080/00018736000101229, Cohen1992}, is screened out by the itinerant electrons.
Thus, understanding the mechanism to overcome this incompatibility between polar instability and metalicity is essential to designing and discovering new polar metals with desired properties.

As the first discovered polar metal, LiOsO$_3$ has been extensively investigated experimentally and theoretically. LiOsO$_3$ shows second-order order-disorder structural phase transition between $R3c$ and $R\bar{3}c$ structures at 140 K~\cite{Shi2013, doi:10.1073/pnas.1908956116, PhysRevB.93.064303, PhysRevResearch.2.033174, PhysRevB.89.201107}. In the high-temperature $R\bar{3}c$ phase, the ferroelectric-like $A_{2u}$ phonon is unstable, which is dominated by the displacements of the Li ions.
The origin of this instability has been discussed from different viewpoints, such as 
instability of Li and O ions~\cite{PhysRevB.90.195113, PhysRevB.90.094108},
short-range geometric and bonding properties~\cite{C5TC03856A, C4RA03946G},
the decoupling electron mechanism~\cite{Laurita2019, Puggioni2014}, 
incomplete screening of the dipole-dipole interaction~\cite{PhysRevB.91.064104},
and hyperferroelectricity~\cite{PhysRevLett.112.127601, Li2016}.
Some other materials have been investigated as well~\cite{doi:10.1021/acs.nanolett.8b00633, Puggioni2014,Fei2018}, but these works focus on individual materials
and systematic investigations have been lacking 
because of the minimal number of known materials.

In addition, doping carriers to ferroelectric insulators has been investigated as an effective way to design a polar metal~\cite{PhysRevLett.109.247601, PhysRevB.94.224107, PhysRevB.97.054107, PhysRevB.86.214103}. Some works predict that the polar atomic displacements are not suppressed and even enhanced by doping in many materials, which are attributed to the so-called meta-screening effect~\cite{PhysRevB.94.224107, PhysRevB.97.054107}. However, they do not consider the finite-temperature effects. Furthermore, they focus on the relations of the ferroelectric insulators and doped ferroelectrics, but the relations between the intrinsic polar metals and these two remain unclear.

\begin{figure}[h]
\vspace{0cm}
\begin{center}
\includegraphics[width=0.5\textwidth]{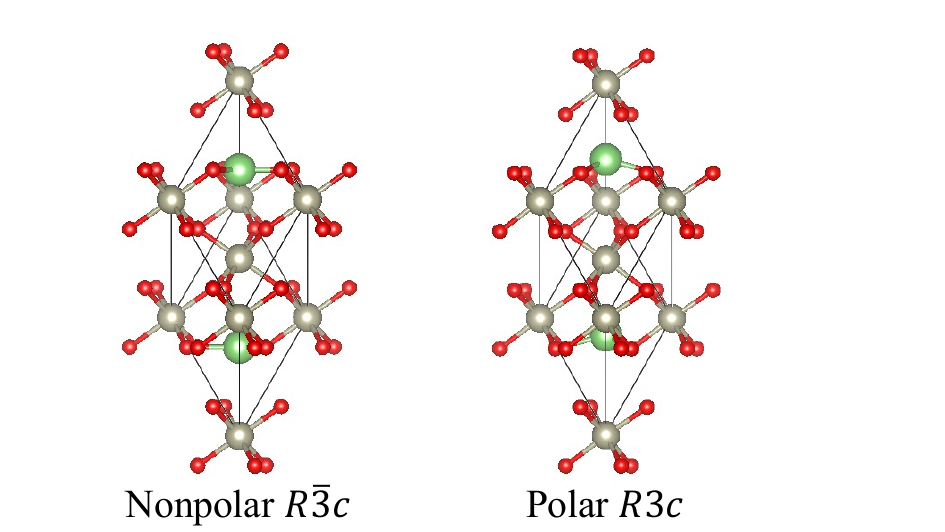}
\caption{
The crystal structure of LiReO$_3$ in the nonpolar $R\bar{3}c$ phase and in the polar $R3c$ phase. The figure of the crystal structures is generated by VESTA~\cite{Momma:db5098}.}
\label{Fig_LiReO3_crystal_structure}
\end{center}
\end{figure}

Recently, LiReO$_3$ have been experimentally revealed to be a polar metal~\cite{Murayama_2023}. LiReO$_3$ is isostructural to LiOsO$_3$ and shows a polar-nonpolar structural phase transition between the high-temperature $R\bar{3}c$ and the low-temperature $R3c$ phases at 170 K, which are shown in Fig.~\ref{Fig_LiReO3_crystal_structure}. 
In addition, LiTaO$_3$, another material in the Li$B$O$_3$ group with a 5$d$ transition metal at the $B$ site, has been reported to be a ferroelectric insulator. Therefore, we consider that Li$B$O$_3$ ($B$ = 5$d$ transition metals) is an ideal platform to perform systematic analysis on the polar metals and to investigate the relations between ferroelectrics, doped ferroelectrics, and polar metals.

In this work, we focus on Li$B$O$_3$ with $B$ = Ta, W, Re, and Os. From the basic electronic-structure calculations, we have confirmed that LiTaO$_3$ is a band insulator while the other three are all metals in both the polar and nonpolar phases. The harmonic phonon calculations show that the $A_{2u}$ mode at $\Gamma$ point has the largest instability in the nonpolar $R\bar{3}c$ phase, which is consistent with the structural phase transitions of LiTaO$_3$, LiReO$_3$, LiOsO$_3$. 

Furthermore, we calculate the $T$-dependence of the crystal structures of these materials based on the self-consistent phonon theory~\cite{PhysRevB.106.224104}. The calculated transition temperatures accurately reproduce the chemical trend. Based on the electronic, phononic, and the structural calculations, we predict that LiWO$_3$ is another polar metal whose $T_c$ is slightly lower than LiReO$_3$ and LiOsO$_3$. The synthesis of LiWO$_3$ has been reported~\cite{inorganics7050063} but its detailed properties has not been measured yet.
Upon doping electrons to the insulating LiTaO$_3$, the high $T_c$ is swiftly but continuously suppressed and approaches those of polar metals. Thus, there is a continuous crossover between the ferroelectric insulator and the polar metals, which are connected by the doped ferroelectrics. Analyzing the interatomic force constants (IFCs) in detail, we explicitly show that the suppression of $T_c$ in polar metals is caused by the screening of the long-range Li-O interactions.

\section{Theory}
We use the structural optimization method based on self-consistent phonon (SCPH) theory~\cite{PhysRevB.106.224104} to calculate the temperature dependence of the crystal structures. SCPH theory is a mean-field theory of the phonon anharmonicity, which has been demonstrated to accurately reproduce finite-temperature properties of strongly anharmonic materials~\cite{doi:10.1080/14786435808243224, PhysRev.165.951, PhysRevLett.129.185901, SOUVATZIS2009888, Monacelli_2021, PhysRevB.105.064112}.

SCPH theory is based on the variational principle of the free energy
\begin{align}
    \mathcal{F} 
  = 
  -k_B T \log \Tr e^{-\beta \hat{\mathcal{H}}_0} + \braket{\hat{H} - \hat{\mathcal{H}}_0}_{\hat{\mathcal{H}}_0}
  \geq F,
  \label{EqVariationalFreeEnergy}
\end{align}
where $F$ is the true free energy, and $\mathcal{F}$ is the variational free energy. $\hat{H}$ is the true Hamiltonian, and $\hat{\mathcal{H}}_0$ is the trial Hamiltonian, which we restrict to be a harmonic Hamiltonian  $
  \hat{\mathcal{H}}_0 = \sum_{\bm{k}\lambda} \hbar \Omega_{\bm{k}\lambda} 
  \Bigl(\hat{a}_{\bm{k}\lambda}^\dag \hat{a}_{\bm{k}\lambda}
  +\frac{1}{2}
  \Bigr)
$ in SCPH theory. The SCPH frequencies $\Omega_{\bm{k}\lambda}$ are considered as the variational parameters, which are adjusted to minimize $\mathcal{F}$~\footnote{We treat the polarization vectors $\epsilon_{\bm{k}\lambda, \alpha\mu}$ as variational parameters as well. However, we explain the case in the fixed-mode approximation for simplicity. Please see Ref.~\cite{PhysRevB.106.224104} for detailed discussion.}. The minimization is performed by solving a self-consistent equation of $\Omega_{\bm{k}\lambda}$~\cite{PhysRevB.92.054301}, which we call the SCPH equation.

In structural optimization, we consider the minimized variational free energy $(\min_{\Omega_{\bm{k}\lambda}}\mathcal{F})$ as the approximate free energy and minimize it with respect to the crystal structures.
We start from the Taylor expansion of the potential energy surface
\begin{align}
\hat{U} = \sum_{n=0}^{\infty} \hat{U}_n,
\end{align}
\begin{align}
&\hat{U}_n \nonumber \\
&=\frac{1}{n!} \sum_{\{\bm{R}\alpha \mu\}} \Phi_{\mu_1 \cdots \mu_n}(\bm{R}_1\alpha_1, \cdots, \bm{R}_n \alpha_n) \hat{u}_{\bm{R}_1 \alpha_1 \mu_1} \cdots \hat{u}_{\bm{R}_n \alpha_n \mu_n}
 \nonumber \\
&=\frac{1}{n!} \frac{1}{N^{n/2-1}} \sum_{\{\bm{k}\lambda\}} \delta_{\bm{k}_1 + \cdots + \bm{k}_n} \widetilde{\Phi} (\bm{k}_1 \lambda_1, \cdots, \bm{k}_n \lambda_n ) \hat{q}_{\bm{k_1} \lambda_1} \cdots \hat{q}_{\bm{k_n }\lambda_n},
  \label{eq_Un}
\end{align}
where $\hat{u}_{\bm{R}\alpha\mu}$ is the $\mu(=x,y,z)$ component of atomic displacement of atom $\alpha$ in the primitive cell at $\bm{R}$, and 
\begin{equation}
\hat{q}_{\bm{k}\lambda} = 
\frac{1}{\sqrt{N}}\sum_{\bm{R}\alpha \mu} 
e^{-i\bm{k}\cdot \bm{R}} 
\epsilon^*_{\bm{k}\lambda,\alpha\mu}\sqrt{M_\alpha} \hat{u}_{\bm{R}\alpha\mu}.
\end{equation}
are those in the normal coordinate representation. $\epsilon_{\bm{k}\lambda, \alpha \mu}$ is the polarization vector of the phonon with mode $\lambda$ and crystal momentum $\bm{k}$.
$M_{\alpha}$ is the mass of the atom $\alpha$.
We call the Taylor expansion coefficients $\Phi_{\mu_1 \cdots \mu_n}(\bm{R}_1\alpha_1, \cdots, \bm{R}_n \alpha_n)$ and $ \widetilde{\Phi} (\bm{k}_1 \lambda_1, \cdots, \bm{k}_n \lambda_n ) $ as interatomic force constants (IFCs).

The IFCs are the functions of atomic positions in the unit cell, which we denote as $X_{\alpha \mu}$ for the $\mu$(=$x,y,z$) component of the atomic position of atom $\alpha$~\footnote{
In Refs.~\cite{PhysRevB.106.224104, PhysRevB.107.134119}, we express the atomic positions by the displacements from the reference positions in the normal coordinate representation, which we write as $q^{(0)}_{\lambda}$. If we denote the atomic coordinates in the reference structure as $X_{\alpha\mu}^{(\text{ref})}$, these quantities are connected by the relation
\begin{align}
q_{\lambda}^{(0)} = \sum_{\alpha \mu} \epsilon_{\bm{0}\lambda, \alpha \mu} \sqrt{M_{\alpha}} (X_{\alpha\mu} - X_{\alpha\mu}^{(\text{ref})}).
\end{align}
}. As the solution of the SCPH equation is determined by the set of IFCs $\widetilde{\Phi}$, we can write down the SCPH free energy as
\begin{align}
F_{\text{SCPH}}(X_{\alpha \mu}) = \mathcal{F}(\widetilde{\Phi}(X_{\alpha \mu}), \Omega_{\bm{k}\lambda}(\widetilde{\Phi}(X_{\alpha\mu}))).
\end{align}
The crystal structure ($X_{\alpha\mu}$)-dependence of the IFCs $\widetilde{\Phi}$ can be calculated using the IFC renormalization~\cite{PhysRevB.106.224104, PhysRevB.107.134119, wallace1972thermodynamics}. Thus, we can calculate the gradient of the SCPH free energy $\frac{\partial F_{\text{SCPH}}(X_{\alpha\mu})}{\partial X_{\alpha \mu}}$ and perform structural optimization with finite-temperature effects. Please see Refs.~\cite{PhysRevB.106.224104, PhysRevB.107.134119} for more details on the IFC renormalization and the structural optimization at finite temperatures.

\section{Simulation Details}

\subsection{Phonon calculations and structural optimizations at finite temperatures}
We use the ALAMODE implementation of the SCPH calculation and SCPH-based structural optimization~\cite{Tadano_2014, PhysRevB.106.224104, PhysRevB.107.134119, PhysRevB.92.054301, PhysRevMaterials.3.033601}. We use the $2\times2\times2$ supercell that contains 80 atoms in the phonon calculations. The lattice constants obtained by structural optimization of VASP are summarized in Table~\ref{table_latconsts_LiBO3}. 

The harmonic IFCs are calculated using the small displacement method with atomic displacements of \SI{0.01}{\angstrom}. The anharmonic IFCs are obtained using the compressive sensing method~\cite{PhysRevLett.113.185501, PhysRevB.92.054301}, which enables efficient extraction of IFCs from a small number of displacement-force data. The displacement-force data is obtained with high-accuracy DFT calculations on the randomly displaced configurations. We use the ab initio molecular dynamics (AIMD) simulation to generate the randomly displaced configurations. We perform the AIMD calculation at 300 K for 16000 steps with the step of 1 fs for LiWO$_3$, LiReO$_3$, and LiOsO$_3$. The first 1000 steps are discarded as thermalization steps, and 300 snapshots are sampled uniformly from the rest 15000 steps. The configurations are generated by adding random atomic displacements of \SI{0.04}{\angstrom} to the 300 AIMD snapshots. The procedure is similar for LiNbO$_3$ and LiTaO$_3$. For these ferroelectric insulators, however, we perform AIMD calculations at 500 K and 750 K for 8000 steps, respectively. In each AIMD calculation, the first 1000 steps are discarded as thermalization steps, and 140 configurations are similarly extracted. Thus, we get 280 displacement-force data for LiNbO$_3$ and LiTaO$_3$, respectively. We choose the calculation settings so that the generated configurations effectively sample the low-energy region of the potential energy surface. Note that high accuracy is not necessary in the AIMD calculations because they are just for generating random structures.

We use $8\times8\times8$ $q$-mesh in SCPH calculations. We fix the shape of the unit cell in SCPH-based structural optimizations because the cell volumes of the considering polar metals do not drastically change on structural phase transitions~\cite{Shi2013, Murayama_2023}. 

\begin{table}[t]
  \caption{The lattice constants that are used in the phonon calculation.}
  \label{table_latconsts_LiBO3}
  \centering
  \begin{ruledtabular}
  \begin{tabular}{ccc}
      Materials & $a$ [\AA] & $c$ [\AA]\\
      \hline
      LiNbO$_3$~\cite{Murayama_2023} & 5.1818  & 13.6313  \\
      LiTaO$_3$ & 5.1885  & 13.6659  \\
      LiWO$_3$ & 5.1744  & 13.5222  \\
      LiReO$_3$~\cite{Murayama_2023} & 5.1267  & 13.3700  \\
      LiOsO$_3$ & 5.1116  & 13.0105  \\
  \end{tabular}
    \end{ruledtabular}
\end{table}

\subsection{DFT calculations}
We employ the \textit{Vienna Ab initio Simulation Package} (VASP)~\cite{PhysRevB.54.11169} for DFT calculations. We use the PBEsol exchange-correlation functional~\cite{PhysRevLett.100.136406} and the PAW pseudopotentials~\cite{PhysRevB.50.17953, PhysRevB.59.1758}. In the high-accuracy calculations, we set the convergence criteria of the SCF loop as $10^{-8}$ eV and the basis cutoff as 600 eV. We use the $4\times4\times4$ Monkhorst-Pack $k$-mesh and accurate precision mode, which suppresses egg-box effects and errors. In the AIMD calculations, we set the convergence criteria of the SCF loop as $10^{-6}$ eV and the basis cutoff as 400 eV. We use the $2\times2\times2$ Monkhorst-Pack $k$-mesh to reduce the computational cost.
In both calculations, we use Gaussian smearing with a width of 0.05 eV. The spin-orbit coupling (SOC) is not included in the anharmonic phonon calculations because it does not affect the low-energy landscape of the potential energy surface~\cite{supplementary}.

\section{Results and Discussion}

\subsection{Electronic Structures of Li$B$O$_3$}
\label{subsec_electronic_structures}

\begin{figure}[h]
\vspace{0cm}
\begin{center}
\includegraphics[width=0.5\textwidth]{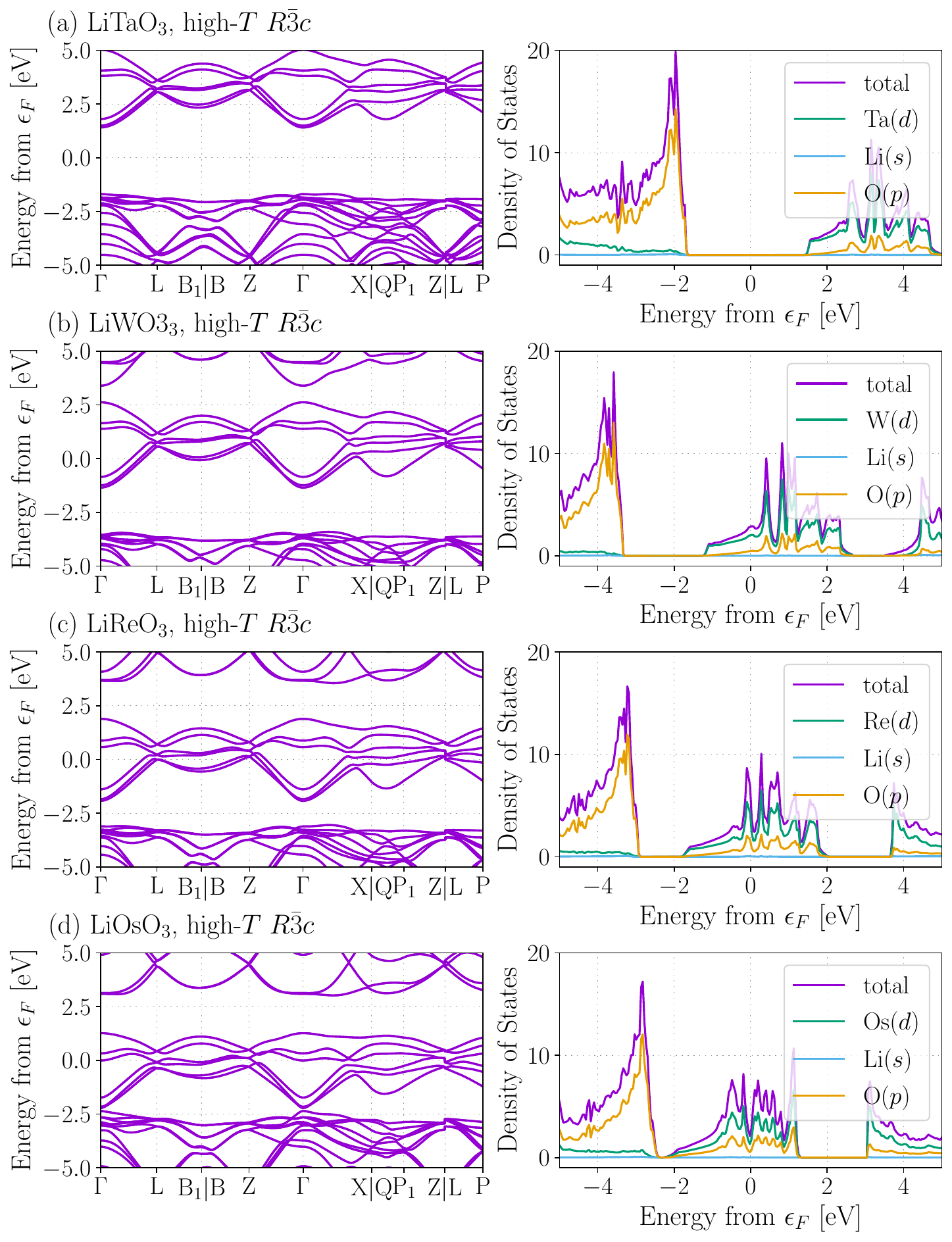}
\caption{
The electronic band structures and density of states (DOS) of (a) LiTaO$_3$, (b) LiWO$_3$, (c) LiReO$_3$, (d) LiOsO$_3$.
The calculations are performed on the high-temperature $R\bar{3}c$ phase, taking into account SOC.}
\label{Fig_LiBO3_highT_withSOC_band_dos}
\end{center}
\end{figure}

We calculate the electronic structures of Li$B$O$_3$ with $B$=Ta, W, Re, Os, which are shown in Fig.~\ref{Fig_LiBO3_highT_withSOC_band_dos}. All the materials show similar band structures with an almost isolated set of 12 bands (including spin degeneracy) near the Fermi level, which consist of hybridized $B$-site $d$-orbitals and O $p$-orbitals.
LiTaO$_3$ is a band insulator as the Fermi level lies in the middle of a band gap. As the number of 5$d$ electrons increases, the Fermi level shifts upward, and the bands become half-filled in LiOsO$_3$. LiWO$_3$, LiReO$_3$, and LiOsO$_3$ are all band metals. 
In the supplementary materials~\cite{supplementary}, we summarize all the calculation results both on the high-$T$ $R\bar{3}c$ and low-$T$ $R3c$ phase, with and without SOC. According to Fig. S1 to S4, LiTaO$_3$ is a band insulator, while LiWO$_3$, LiReO$_3$, and LiOsO$_3$ are metals in the low-$T$ $R3c$ phase as well. Thus, the metal-insulator transitions do not occur with the structural phase transitions in the target materials. This is consistent with the experimental observation that LiReO$_3$ is a polar metal~\cite{Murayama_2023}, and supports our prediction that LiWO$_3$ is another polar metal, which we discuss later.

In addition, we would like to add a short discussion on electronic correlations. The target materials Li$B$O$_3$ can be strongly correlated electron systems due to the 5$d$ transition metals in the $B$ site. In particular, electronic correlations of LiOsO$_3$ have been shown to be essential to precisely describe the electronic properties~\cite{PhysRevB.104.115130, PhysRevB.93.161113, PhysRevMaterials.4.045001}.
However, we do not explicitly consider the electronic correlations in this work. This is because the effect of electronic correlation will be the largest in LiOsO$_3$, which have half-filled $t_{2g}$ bands, but it is not strong enough to cause metal-insulator transition. Thus, the conventional DFT calculations should be accurate enough to discuss distinctions between metals and insulators and basic electronic structures. Furthermore, previous works show that the electronic correlation little affects the structural properties~\cite{PhysRevB.90.195113}, which is the main motivation of this work.

\subsection{Harmonic phonons of Li$B$O$_3$}
\label{subsec_harmonic_phonons}

\begin{figure}[h]
\vspace{0cm}
\begin{center}
\includegraphics[width=0.5\textwidth]{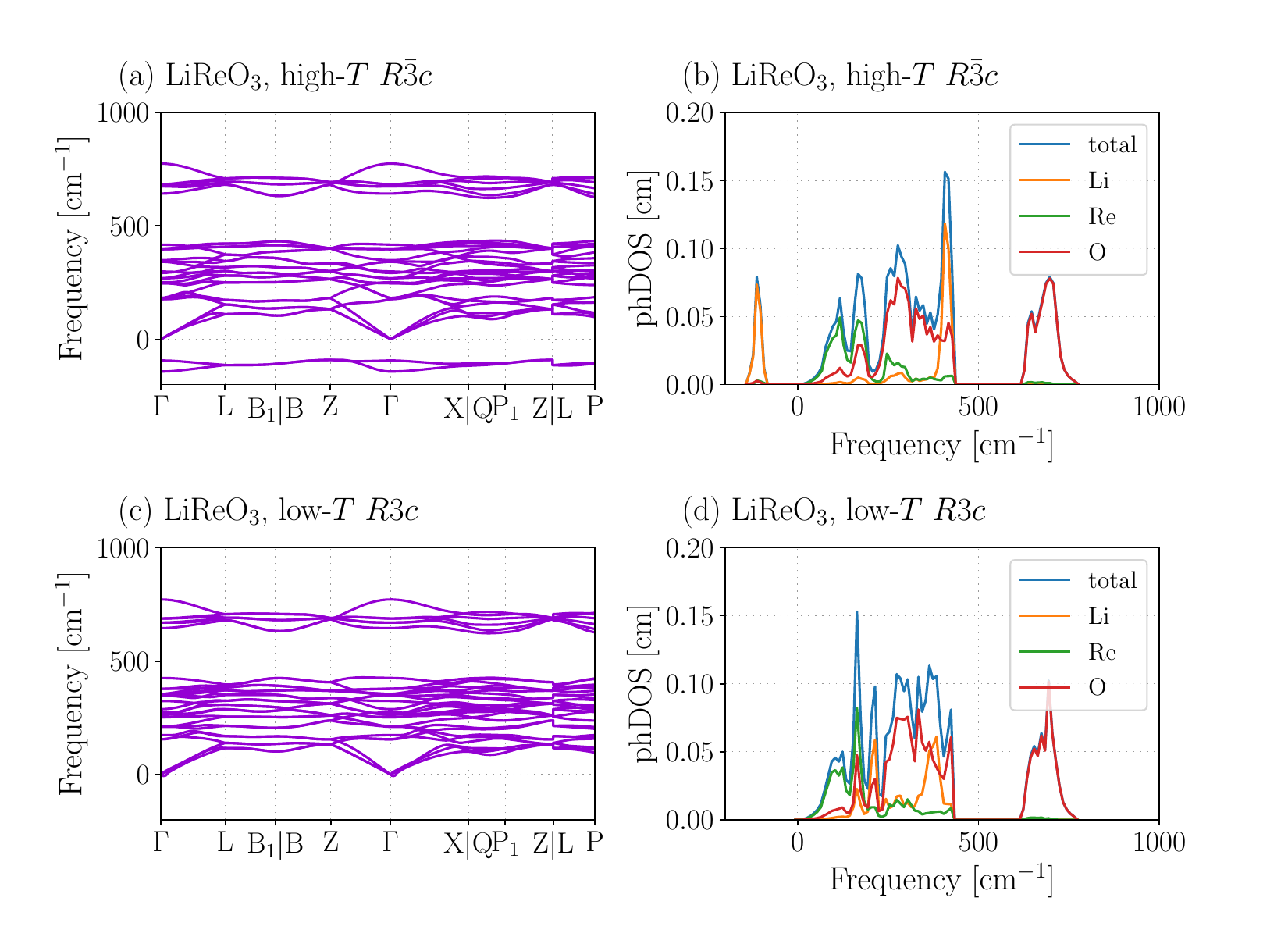}
\caption{
The harmonic phonon dispersion and atom-projected density of states of LiReO$_3$ in the high-$T$ $R\bar{3}c$ phase and in the low-$T$ $R3c$ phase.}
\label{Fig_LiReO3_phband_phdos}
\end{center}
\end{figure}

We calculate the harmonic phonon dispersions and atom-projected density of states of Li$B$O$_3$. Note that we neglect SOC from this section because SOC hardly affects the low-energy region of the potential energy surface, which we show in Sec. II in the supplementary materials~\cite{supplementary}. Figure ~\ref{Fig_LiReO3_phband_phdos} 
shows the calculation results of LiReO$_3$.
LiReO$_3$ has a pair of unstable modes in the high-$T$ $R\bar{3}c$ phase, which is dominated by Li ions. The most unstable mode is the ferroelectric-like $A_{2u}$ mode at $\Gamma$ point that causes the transition to the low-$T$ $R3c$ phase, which is common to all $B$ = Ta, W, and Os cases~\cite{supplementary}. The non-dispersive nature of the soft modes suggests that the transition is caused by the on-site instability of the loosely bonded small Li ions, as suggested for LiOsO$_3$~\cite{PhysRevB.89.201107, PhysRevB.90.094108}. The imaginary phonon is lifted in the low-temperature phase, and the instability disappears.

In Section~\ref{subsec_electronic_structures}, we saw that the electronic bands of Li$B$O$_3$ near the Fermi level are dominated by the $B$-site $d$ orbitals and O $p$ orbitals, while the contribution of Li ions, which dominates the instability of the high-symmetry phase, is negligible. Thus, LiReO$_3$ seems consistent with the decoupling electron mechanism~\cite{Puggioni2014}, that the electronic state near the Fermi level and the polar atomic displacements are decoupled, as discussed on LiOsO$_3$~\cite{Laurita2019, PhysRevB.90.094108}.

\subsection{Structural phsae transitions of Li$B$O$_3$}
\begin{figure}[h]
\vspace{0cm}
\begin{center}
\includegraphics[width=0.5\textwidth]{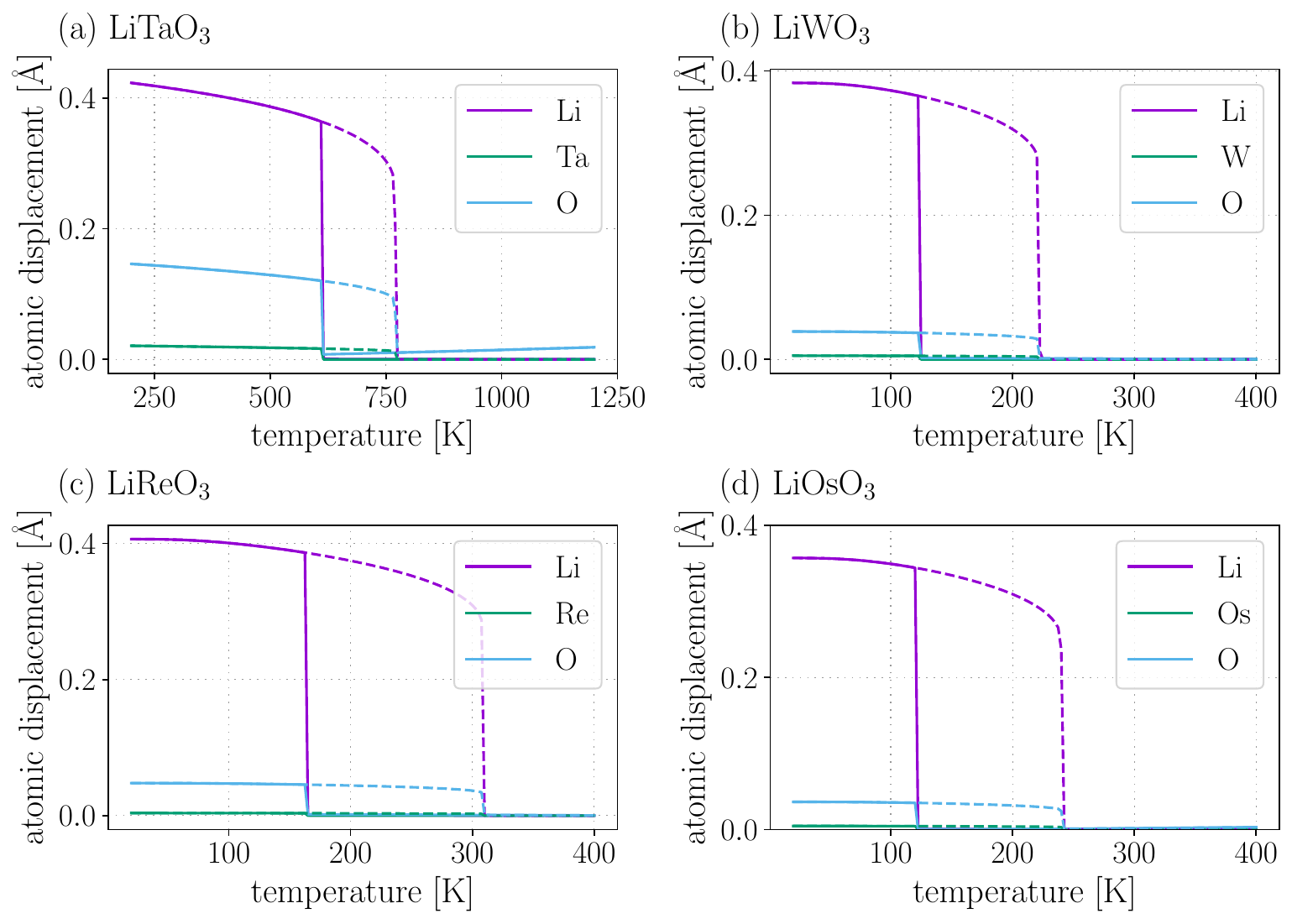}
\caption{
Temperature dependence of the atomic displacements of Li$B$O$_3$ with $B$ = (a) Ta, (b) W, (c) Re, (d) Os. The solid lines show the cooling calculations, in which we start from the high-temperature and add a slight atomic displacement along the unstable $A_{2u}$ mode until the transition to the low-temperature phase is induced. The dashed lines show the heating calculations, in which we start from the low-temperature $R3c$ phase.}
\label{Fig_LiBO3_T_atom_disp}
\end{center}
\end{figure}

Based on the above discussions, we apply the SCPH-based structural optimization at finite temperatures to Li$B$O$_3$ with $B$ = Ta, W, Re, Os. The temperature-dependence of the atomic displacements are shown in Fig.~\ref{Fig_LiBO3_T_atom_disp}. The displacements of Li and $B$-site ions are zero at high temperatures, while they are finite at low temperatures. Hence, the polar-nonpolar structural phase transitions of these materials are reproduced by theoretical calculations. The transition temperatures, which are estimated from the crossing points of the SCPH free energies, are summarized in Table~\ref{table_LiBO3_Tc}. We can see that the calculated $T_c$ of each material is compatible with the experimental values. In addition, the calculation results reproduce the chemical trend, i.e., $T_c$ of the ferroelectric insulators ($B$=Nb, Ta) are much higher than those of polar metals ($B$=Re, Os). The chemical trends within each class (the ferroelectrics and the polar metals) are also accurately reproduced. This success further demonstrates the accuracy of our recently developed method~\cite{PhysRevB.106.224104}.

As shown in Fig. S7 in the supplementary materials~\cite{supplementary},
the ferroelectric-like $A_{2u}$ mode that drives the polar-nonpolar structural phase transition has the largest instability in the high-temperature $R\bar{3}c$ phase of LiWO$_3$.
 In addition, LiWO$_3$ is metallic in both $R\bar{3}c$ and $R3c$ phases, which we can see from Fig. S2. Thus, we predict that LiWO$_3$ is another isostructural polar metal whose transition temperature $T_c$ is slightly lower than LiReO$_3$ and LiOsO$_3$.

\begin{figure}[h]
\vspace{0cm}
\begin{center}
\includegraphics[width=0.5\textwidth]{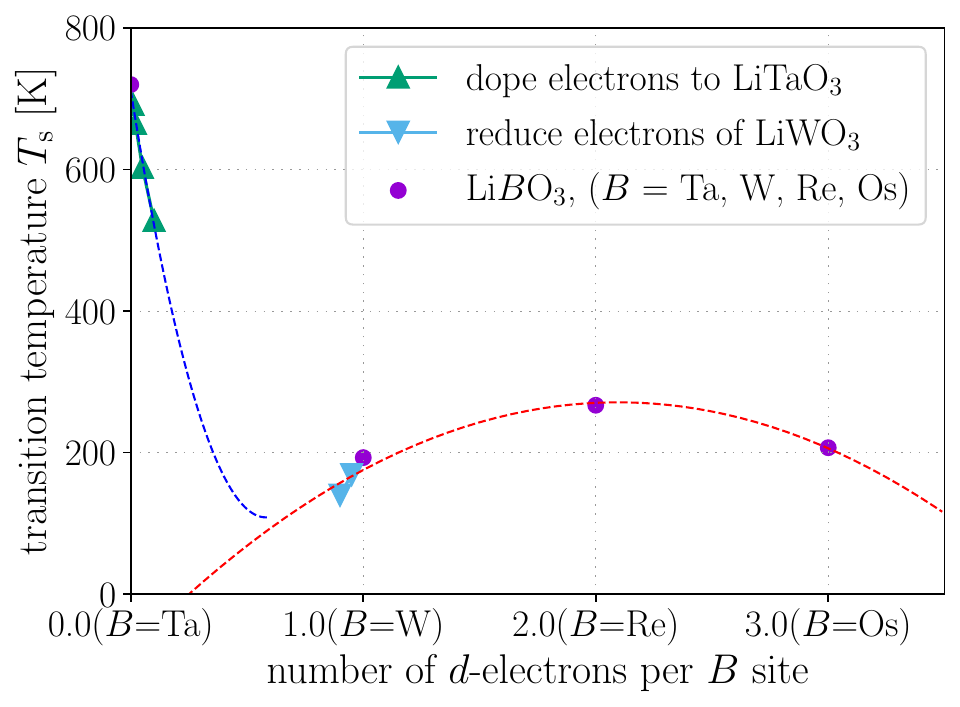}
\caption{
The transition temperatures of Li$B$O$_3$ with different numbers of electrons per $B$ site. The dotted lines are guide to the eye.}
\label{Fig_Bsite_Tc}
\end{center}
\end{figure}

As depicted in Fig.~\ref{Fig_Bsite_Tc}, the calculated $T_c$ of the polar metals Li$B$O$_3$ ($B$ = W, Re, Os) have a dome-like dependence on the number of $d$ electrons per $B$ site, while LiTaO$_3$ significantly deviates from this trend.
Here, we focus on the region between the ferroelectric insulator and the polar metals. We dope electrons to the insulating LiTaO$_3$ by changing the number of electrons in DFT calculations and calculate the change of $T_c$. From Fig.~\ref{Fig_Bsite_Tc}, the $T_c$ of LiTaO$_3$ is quickly suppressed and approaches those of polar metals when electrons are doped.
Note that such calculations to change the number of electrons are not necessarily accurate because of the uniform positive background. In fact, we fix the shape of the unit cell when we change the number of electrons because the lattice constants optimized by DFT calculations largely deviate from the chemical trend.
However, we consider the quick suppression of $T_c$ in doped LiTaO$_3$ is qualitatively correct because $T_c$ changes much more slowly and follows the dome-like trend when we reduce the number of electrons in LiWO$_3$, as shown in Fig.~\ref{Fig_Bsite_Tc}.

Because the change of $T_c$ is continuous, we consider that there is a continuous crossover between the ferroelectric insulators and the polar metals, which are connected by doped ferroelectrics.
Upon doping electrons to ferroelectric insulators, the polar instabilities is suppressed at finite temperatures, even when the polar displacements at zero temperature remain unchanged.

\begin{table}[t]
  \caption{The calculated and experimental transition temperatures ($T_c$) of Li$B$O$_3$ with $B$ = Nb, Ta, W, Re, Os.}
  \label{table_LiBO3_Tc}
  \centering
  \begin{ruledtabular}
  
  \begin{tabular}{ccc}
      Materials & calculation [K] & experiment [K]\\
      \hline
      LiNbO$_3$ & 1350~\cite{Murayama_2023}  & 1480~\cite{PhysRevB.53.1193}  \\
      LiTaO$_3$ & 723  & 950~\cite{PhysRevB.53.1193}  \\
      LiWO$_3$ & 193  & -  \\
      LiReO$_3$ & 267~\cite{Murayama_2023}  & 170~\cite{Murayama_2023}  \\
      LiOsO$_3$ & 207  & 140~\cite{Shi2013}  \\
  \end{tabular}
    \end{ruledtabular}
\end{table}

\subsection{Chemical trend of $T_c$ in the polar metals Li$B$O$_3$}

In this section, We discuss the origin of the dome-like structure of $T_c$ of polar metals in Fig.~\ref{Fig_Bsite_Tc}. Up to the quartic order, the double well potential along the soft mode can be written as
\begin{align}
U(q_{\lambda}) 
&= \frac{1}{2} \widetilde{\Phi}(\bm{0}\lambda, \bm{0}\lambda) q_{\lambda}^2
+ \frac{1}{4!} \widetilde{\Phi}(\bm{0}\lambda, \bm{0}\lambda, \bm{0}\lambda, \bm{0}\lambda)q_{\lambda}^4
\\&=
\frac{1}{2} \widetilde{\Phi}_2 q_{\lambda}^2
+ \frac{1}{4!} \widetilde{\Phi}_4 q_{\lambda}^4,
\end{align}
where we defined $\widetilde{\Phi}_2 = \widetilde{\Phi}(\bm{0}\lambda, \bm{0}\lambda)$ and $\widetilde{\Phi}_4 = \widetilde{\Phi}(\bm{0}\lambda, \bm{0}\lambda, \bm{0}\lambda, \bm{0}\lambda)$ for notational simplicity. 
$q_{\lambda}$ is the atomic displacement along the ferroelectric-like soft mode in normal coordinate representation.
The depth of this double well potential is $\frac{3\widetilde{\Phi}^2_2}{2\widetilde{\Phi}_4}$, which is roughly proportional to $T_c$ as shown in Table~\ref{table_doublewelldepth_Tc}. In Table~\ref{table_doublewelldepth_Tc}, $T_{c,\text{est}}$ are the transition temperatures estimated from the assumption that $T_c \propto \frac{3\widetilde{\Phi}^2_2}{2\widetilde{\Phi}_4}$ and $T_{c}$ of LiReO$_3$ is 267 K. Since $T_{c,\text{est}}$ reproduces the trend of the calculated transition temperatures $T_{c,\text{calc}}$, we can conclude that the assumption $T_c \propto \frac{3\widetilde{\Phi}^2_2}{2\widetilde{\Phi}_4}$ holds approximately. 
Thus, the chemical trend of $T_c$ of Li$B$O$_3$ can be explained by the change of $\widetilde{\Phi}_2$ and $\widetilde{\Phi}_4$ among these materials.

According to Table~\ref{table_doublewelldepth_Tc}, $\widetilde{\Phi}_4$ monotonically increases from $B$=Ta to $B$=Os. This is presumably because the quartic interaction arises from the short-range repulsive forces between ions, which generally get larger as the lattice constants get smaller.
$|\widetilde{\Phi}_2|$ takes the largest value in LiTaO$_3$, which leads to the highest $T_c$ among the target materials. In the polar metals, $|\widetilde{\Phi}_2|$ also monotonically increases from $B$=W to $B$=Os.
If the instability originates from the competition of the short-range repulsion and the long-range dipole-dipole interaction, which is usually the case in ferroelectric insulators, the chemical trend of $\widetilde{\Phi}_2$ would be opposite because the short-range repulsion quickly gets prominent when the lattice constant shrinks. Thus, this $B$ site-dependence of $|\widetilde{\Phi}_2|$ supports that the instability of the polar metals has a short-range origin, which has been suggested for LiOsO$_3$ in previous research~\cite{C5TC03856A}.

\begin{table*}[t]
  \caption{
  The interatomic force constants along the ferroelectric-like soft mode $\bm{0}\lambda$ and the estimated transition temperatures. $\widetilde{\Phi}_2$ and $\widetilde{\Phi}_4$ are the IFCs along the soft mode, which are defined as 
  $\widetilde{\Phi}_2 = \widetilde{\Phi}(\bm{0}\lambda, \bm{0}\lambda)$ and $\widetilde{\Phi}_4 = \widetilde{\Phi}(\bm{0}\lambda, \bm{0}\lambda, \bm{0}\lambda, \bm{0}\lambda)$. $T_{c,\text{est}}$ are the transition temperatures estimated from the assumption that $T_c \propto \frac{3\widetilde{\Phi}^2_2}{2\widetilde{\Phi}_4}$ and $T_{c}$ of LiReO$_3$ is 267 K. $T_{c,\text{calc}}$ is the $T_c$ calculated by the SCPH-based structural optimization, which are also summarized in Table~\ref{table_LiBO3_Tc}.}
  \label{table_doublewelldepth_Tc}
  \centering
  \begin{ruledtabular}
  
  \begin{tabular}{ccccc}
  &
  $\widetilde{\Phi}_2$ [Ry/($a_B^2$ amu)]& 
  $\widetilde{\Phi}_4$ [Ry/($a_B^2$ amu)$^2$] & 
  $T_{c, \text{est}}$ [K] & $T_{c, \text{calc}}$ [K]\\
  \hline
LiTaO$_3$  &  -1.963 $\times10^{-3}$ &  0.706 $\times10^{-3}$ &  586  &  723 \\
LiWO$_3$  &  -1.091 $\times10^{-3}$ &  0.741 $\times10^{-3}$ &  173  &  193 \\
LiReO$_3$  &  -1.541 $\times10^{-3}$ &  0.955 $\times10^{-3}$ &  267  &  267 \\
LiOsO$_3$  &  -1.591 $\times10^{-3}$ &  1.326 $\times10^{-3}$ &  205  &  207 \\
  \end{tabular}
    \end{ruledtabular}
\end{table*}

\subsection{Difference between ferroelectric insulators and polar metals}

\begin{figure}[h]
\vspace{0cm}
\begin{center}
\includegraphics[width=0.5\textwidth]{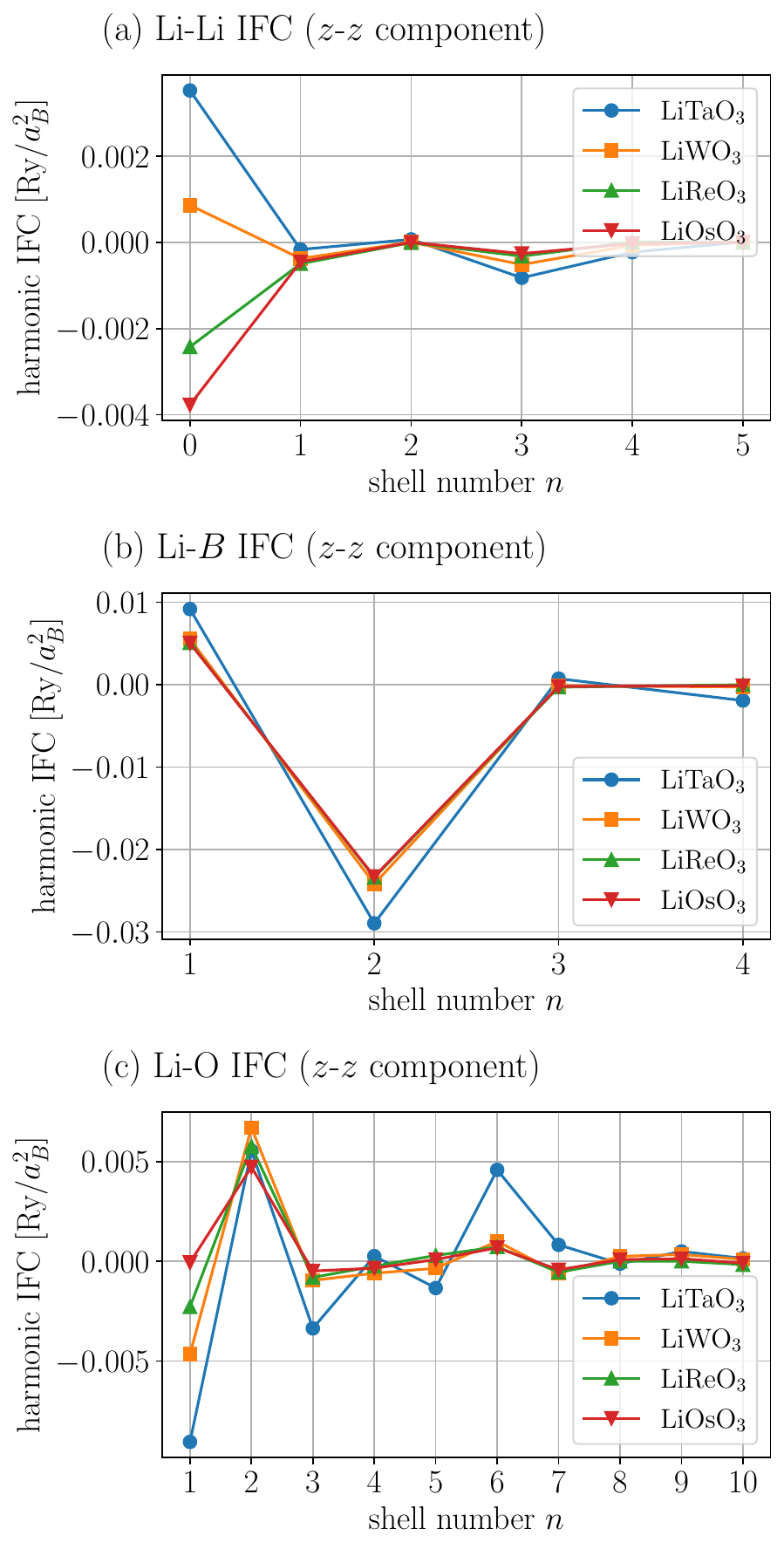}
\caption{
The $zz$ components of the interatomic force constants of the $n$-th nearest neighbor shells from a Li ion of Li$B$O$_3$ ($B$ = Ta, W, Re, Os).
The $n$-th nearest neighbor shells of the element pair Li-$A$ is defined as the $n$-th nearest group of $A$ atoms when we fix a Li atom as the center and classify the $A$ atoms around it in terms of the distance from the fixed Li atom. The plotted IFCs of the $n$-th nearest neighbor shell of the element pair Li-$A$ is $\Phi_{zz}(\text{center Li atom}, \bm{R}\alpha)$, where $\bm{R}\alpha$ is included in the nearest neighbor shell.
Note that we consider the onsite IFC as the contribution of the zero-th nearest neighbor.
}
\label{Fig_LiBO3_shell_IFC_zz}
\end{center}
\end{figure}

Lastly, we discuss the origin of the difference in the transition temperatures between the ferroelectric insulator LiTaO$_3$ and the polar metals Li$B$O$_3$ ($B$ = W, Re, Os). In Fig.~\ref{Fig_LiBO3_shell_IFC_zz}, we plot the $z$-$z$ components of the interatomic force constants (IFCs) of the $n$-th nearest neighbor (n.n.) shells from a Li ion. Note that similar discussions can be done on other components of the IFCs, which is shown in Section IV in the supplementary materials~\cite{supplementary}.
The IFCs of different element pairs are plotted separately.
The $n$-th nearest neighbor shell of the element pair Li-$A$ is defined as the $n$-th nearest group of $A$ atoms when we fix a Li atom and classify the $A$ atoms around it in terms of the distance from the fixed Li atom. The atomic distances of these $n$-th nearest neighbor shells of LiReO$_3$ are summarized in Table~\ref{table_shell_LiReO3}. The cases of $B$ = Ta, W, Os are shown in Tables S2 to S4 in the supplementary materials~\cite{supplementary}, which are almost the same as the LiReO$_3$ case. Since the soft modes of Li$B$O$_3$ are dominated by Li and O displacements, let us consider the results of the element pairs Li-Li and Li-O [Fig.~\ref{Fig_LiBO3_shell_IFC_zz} (a), (c)]. 
In fact, the contributions of the Li-$B$ IFCs to the potential energy surface along the soft modes are small as we later discuss in this section.
The short-range IFCs with atomic distances smaller than \SI{3.0}{\angstrom} [0-th n.n. shell of Li-Li (Li onsite) and first and second n.n. shell of Li-O] are finite in both the ferroelectric insulator and the polar metals. These IFCs seem to change almost linearly when the $B$-site ion is changed.
On the other hand, the long-range IFCs, which we define as IFCs with atomic distances larger than \SI{3.0}{\angstrom} in this paper
[e.g. 3rd n.n. shell of Li-Li, 3rd, 5th, and 6th n.n. shells of Li-O], are finite in LiTaO$_3$ but are close to zero in the polar metals. 

\begin{figure}[h]
\vspace{0cm}
\begin{center}
\includegraphics[width=0.5\textwidth]{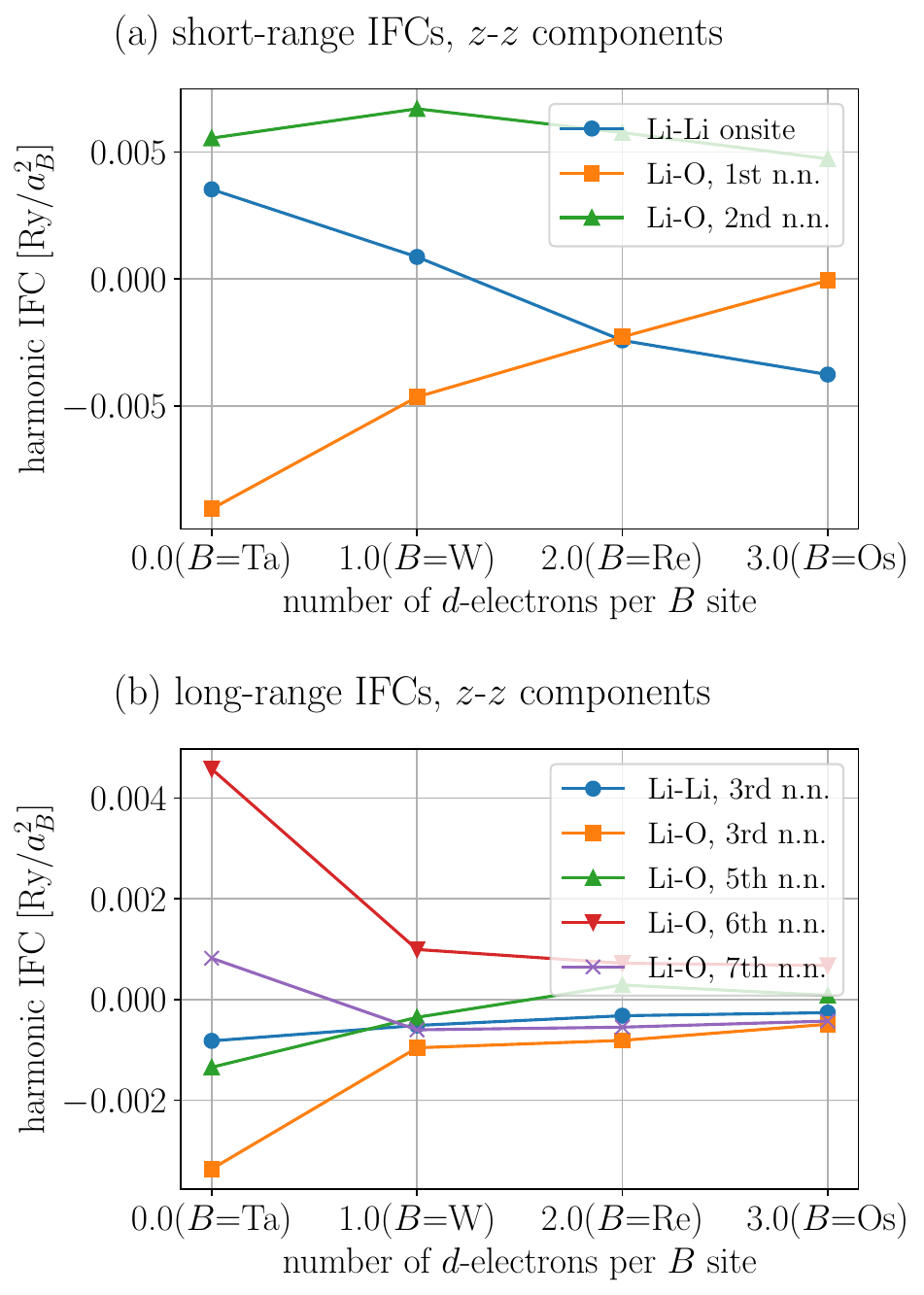}
\caption{
$B$-site dependence of the $z$-$z$ components of the interatomic force constants (IFCs) of $n$-th nearest neighbor (n.n.) shells of different element pairs. (a) The short-range IFCs with atomic distance smaller than 3 \AA. (b) The long-range IFCs with atomic distance larger than 3 \AA.
}
\label{Fig_LiBO3_B_vs_IFC_zz}
\end{center}
\end{figure}

This distinction between the short-range and long-range IFCs can be seen more clearly in Fig.~\ref{Fig_LiBO3_B_vs_IFC_zz}, in which we plot the $B$-dependence of the IFCs. As we see from Fig.~\ref{Fig_LiBO3_B_vs_IFC_zz} (a), the short-range IFCs show systematic dependence on $B$ site ions, and there is no apparent difference between the ferroelectric insulator and the polar metals. Figure ~\ref{Fig_LiBO3_B_vs_IFC_zz} (b) shows representative examples of middle-to-long-range IFCs that have relatively large contributions. These long-range IFCs have a significant finite value in LiTaO$_3$, but they are small and almost constant in polar metals, which can ascribed to the screening of the Coulombic interaction by the itinerant electrons. Although we discuss the importance of O-O interactions in the supplementary materials~\cite{supplementary}, we do not consider them here because it is difficult to choose important components such as $z$-$z$ components. As the O-sites have lower symmetry, their displacements in the soft mode are not along the $z$ direction, and atoms that belong to the same nearest neighbor shell can be displaced in different directions.

\begin{table}[t]
  \caption{$n$-th nearest neighbor shells from a Li ion of LiReO$_3$. The shell number $n$ with the number of atoms in the $n$-th n.n. shell, and the corresponding atomic distances. The calculation is performed on the crystal structure in the high-temperature $R\bar{3}c$ phase without SOC. Note that the number of atoms in the shells are calculated within the 2$\times$2$\times$2 supercell considering the periodic boundary condition.}
  \label{table_shell_LiReO3}
  \centering
  \begin{ruledtabular}
  
  \begin{tabular}{ccc}
\multicolumn{3}{c}{Li-Li} \\
shell number $n$ & num. of atoms & distance [\AA] \\ \hline
0  &  1  &  0.0000 \\
1  &  6  &  3.7049 \\
2  &  3  &  5.1267 \\
3  &  3  &  5.3500 \\
4  &  2  &  6.3253 \\
5  &  1  &  7.4098 \\
\hline
\multicolumn{3}{c}{Li-$B$} \\
shell number $n$ & num. of atoms & distance [\AA] \\ \hline
1  &  6  &  3.1626 \\
2  &  2  &  3.3425 \\
3  &  2  &  6.0237 \\
4  &  6  &  6.1200 \\
\hline
\multicolumn{3}{c}{Li-O} \\
shell number $n$ & num. of atoms & distance [\AA] \\ \hline
1  &  3  &  1.9852 \\
2  &  6  &  2.7368 \\
3  &  3  &  3.1414 \\
4  &  6  &  4.2033 \\
5  &  6  &  4.4773 \\
6  &  6  &  4.7314 \\
7  &  6  &  4.8574 \\
8  &  6  &  5.2769 \\
9  &  3  &  5.7065 \\
10  &  3  &  6.2041 \\
  \end{tabular}
    \end{ruledtabular}
\end{table}

\begin{figure}[h]
\vspace{0cm}
\begin{center}
\includegraphics[width=0.5\textwidth]{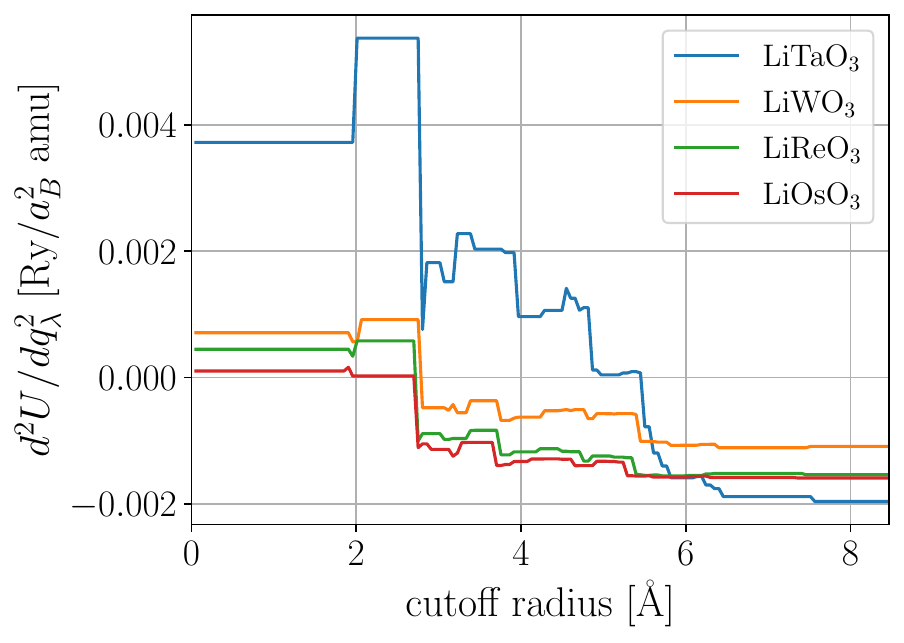}
\caption{
Cutoff-dependence of the curvature of the potential energy surface $\frac{d^2U}{dq_{\lambda}^2}$. 
We set the harmonic IFCs with the atomic distance larger than the cutoff as zero and calculate the curvature 
$
\frac{d^2U}{dq_{\lambda}^2} = \sum'_{\alpha, \bm{R}_1\alpha_1, \mu \mu_1} 
\frac{\epsilon_{\bm{0} \lambda,\alpha \mu}}{\sqrt{M_{\alpha}}}
\frac{\epsilon_{\bm{0} \lambda,\alpha_1 \mu_1}}{\sqrt{M_{\alpha_1}}}
\Phi_{\mu \mu_1}(\bm{0}\alpha, \bm{R}_{1} \alpha_{1}).
$, where $\sum'$ is the sum restricted to the IFCs with atomic distance smaller than the cutoff radius.
$q_{\lambda}$ is the atomic displacement along the ferroelectric-like soft mode in normal-coordinate representation. The calculation is performed on the nonpolar $R\bar{3}c$ phase without SOC.}
\label{Fig_LiBO3_d2UdQ2_along_softmodes}
\end{center}
\end{figure}

Furthermore, we apply cutoffs to the harmonic IFCs and investigate the change of instabilities along the ferroelectric-like soft modes. Figure ~\ref{Fig_LiBO3_d2UdQ2_along_softmodes} shows the cutoff dependence of the curvature of the potential energy surface. Here, we set the harmonic IFCs with atomic distance larger than the cutoff as zero and calculate 
\begin{align}
\frac{d^2U}{dq_{\lambda}^2} = \sum_{\text{dist}(\bm{0}\alpha, \bm{R}_1\alpha_1) < \text{cutoff}} 
\sum_{\mu \mu_1}
\frac{\epsilon_{\bm{0} \lambda,\alpha \mu}}{\sqrt{M_{\alpha}}}
\frac{\epsilon_{\bm{0} \lambda,\alpha_1 \mu_1}}{\sqrt{M_{\alpha_1}}}
\Phi_{\mu \mu_1}(\bm{0}\alpha, \bm{R}_{1} \alpha_{1}).
\end{align}
$q_{\lambda}$ is the atomic displacement in normal coordinate representation along the soft mode, and the polarization vector of the soft mode $\epsilon_{\bm{k}\lambda, \alpha\mu}$ is fixed when the dynamical matrix is altered by applying the cutoff. 
In Fig.~\ref{Fig_LiBO3_d2UdQ2_along_softmodes}, the polar metals with $B$ = W, Re, Os show similar hehaviors. The instability ($\frac{d^2U}{dq_{\lambda}^2} < 0$) appears when the cutoff radius is around 3.0 \AA~and the curvature is almost constant when the cutoff is large. Thus, the polar instabilities of the polar metals are dominated by the short-range IFCs, and the contributions from the long-range IFCs are relatively small, consistent with the discussion in the last paragraph. On the other hand, the long-range IFCs have considerable contributions in LiTaO$_3$, where the largest instability appears only when the long-range IFCs are considered. 

\begin{figure*}[ht]
\vspace{0cm}
\begin{center}
\includegraphics[width=1.0\textwidth]{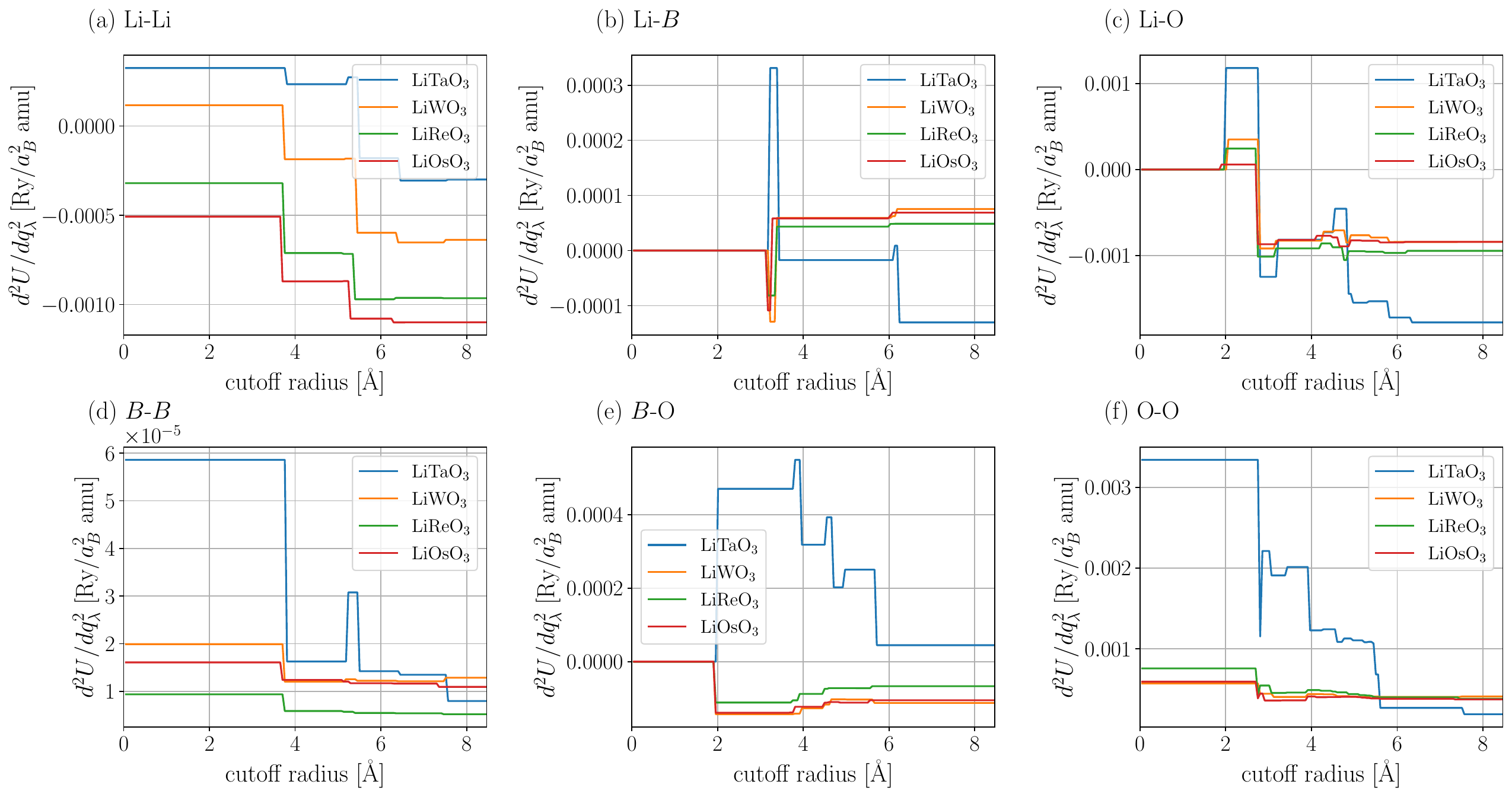}
\caption{
Contributions from IFCs of different element pairs to the curvature of the potential energy surface $\frac{d^2U}{dq_{\lambda}^2}$. 
To calculate the cutoff radius-dependence, we set the harmonic IFCs with the atomic distance larger than the cutoff as zero and calculate the curvature $
\frac{d^2U}{dq_{\lambda}^2} = \sum'_{\alpha, \bm{R}_1\alpha_1, \mu \mu_1} 
\frac{\epsilon_{\bm{0} \lambda,\alpha \mu}}{\sqrt{M_{\alpha}}}
\frac{\epsilon_{\bm{0} \lambda,\alpha_1 \mu_1}}{\sqrt{M_{\alpha_1}}}
\Phi_{\mu \mu_1}(\bm{0}\alpha, \bm{R}_{1} \alpha_{1}).
$, where $\sum'$ is the sum restricted to the IFCs with atomic distance smaller than the cutoff radius.
$q_{\lambda}$ is the atomic displacement along the ferroelectric-like soft mode in normal-coordinate representation. The calculation is performed on the nonpolar $R\bar{3}c$ phase without SOC.
}
\label{Fig_LiBO3_d2UdQ2_elempairs}
\end{center}
\end{figure*}

In Fig.~\ref{Fig_LiBO3_d2UdQ2_elempairs}, we separate the result of Fig.~\ref{Fig_LiBO3_d2UdQ2_along_softmodes} to contributions from different element pairs. The contributions of the element pairs that contain $B$-site ions are small because the displacements of $B$-site ions are negligibly small in the soft modes of these materials as shown in Table S1 in the supplementary materials~\cite{supplementary}. The contribution of Li-Li IFCs [Fig.~\ref{Fig_LiBO3_d2UdQ2_elempairs} (a)] shows similar hehaviors in all materials, and the instability gets monotonically more significant from $B$=Ta to $B$=Os. This is consistent with the $B$-site dependence of the onsite Li IFC that the Li site gets more unstable from $B$=Ta to $B$=Os, which is shown in Fig.~\ref{Fig_LiBO3_shell_IFC_zz}~(a) and Fig.~\ref{Fig_LiBO3_B_vs_IFC_zz}~(a). As shown in Fig.~\ref{Fig_LiBO3_d2UdQ2_elempairs} (c), the Li-O interactions are crucial to the largest instability and the highest $T_c$ of LiTaO$_3$ among the target materials, to which the middle-to-long range IFCs have a significant contribution.

The short-range part of  Fig.~\ref{Fig_LiBO3_d2UdQ2_along_softmodes} shows clear difference between LiTaO$_3$ and the polar metals although the $B$-site dependence of the short-range IFCs is rather systematic as shown in Fig.~\ref{Fig_LiBO3_B_vs_IFC_zz}  (a). These seemingly contradicting results can be ascribed to the difference in the polarization vector. In Section V in the supplementary materials~\cite{supplementary}, we discuss that larger O contribution to the soft mode is essential to the large instability of LiTaO$_3$, for which the long-range Li-O and O-O interactions play an important role.

\subsection{Discussion on the origin of structural phase transitions in polar metals}
Here, we briefly discuss our understanding on the structural phase transitions of the polar metals in comparison with the previous works on LiOsO$_3$.
We have clarified that the drastic suppression of $T_c$ from the ferroelectric insulator LiTaO$_3$ to the polar metals are caused by the screening of the long-range IFCs, which is ascribed to the itinerant electrons. The polar instability of the polar metals originate from the short-range IFCs, which show linear-like dependence on $B$-site atomic numbers through LiTaO$_3$ to LiWO$_3$. 
These results support that the polar metals LiReO$_3$ and LiOsO$_3$ have short-range origins as suggested in Ref.~\cite{PhysRevB.90.094108, C5TC03856A, C4RA03946G}. We cannot distinguish the chemical and geometrical effect because we have not investigated the nature of chemical bondings around the Li ions. 

The soft modes in these polar metals are dominated by the Li ions, while the electronic structures near the Fermi level consist of the O and $B$-site orbitals, which seems consistent with the decoupling electron mechanism. However, the weak coupling between the soft mode and the low-energy electronic structure does not prevent the screening of the long-range interactions, as suggested by Anderson and Blount~\cite{PhysRevLett.14.217}. Indeed, our calculations show that the unscreened interactions are not the main driving force of the instability of the high-symmetry $R\bar{3}c$ phases of the polar metals. However, this does not necessarily exclude the possibility that the remaining weak off-site interactions contribute to the emergence of the long-range order below $T_c$~\cite{PhysRevB.91.064104}.

\section{Conclusions}
We perform a systematic analysis on Li$B$O$_3$ with $B$ = Ta, W, Re, Os. LiTaO$_3$ is a ferroelectric insulator, while LiReO$_3$ and LiOsO$_3$ are polar metals. The DFT calculations show that LiTaO$_3$ is a band insulator while the other three are metals, consistent with the experiments. The phonon calculations show that the ferroelectric-like $A_{2u}$ mode has the largest instability in the high-temperature $R\bar{3}c$ phase, consistent with the structural phase transitions. We then apply the SCPH-based structural optimization to Li$B$O$_3$ and accurately reproduce the chemical trend of the transition temperatures. From these calculations, we predict that LiWO$_3$ is another polar metal yet to be tested experimentally. In the end, we perform a detailed analysis on the interatomic force constants and explicitly show that the suppression of $T_c$ in polar metals can be ascribed to the screening of the long-range IFCs caused by the itinerant electrons.

\begin{acknowledgments}
This work was supported by JSPS KAKENHI Grant Number 21K03424 and 19H05825, Grant-in-Aid for JSPS Fellows (No. 22KJ1028), and JST-PRESTO (JPMJPR20L7 and JPMJPR23J6).
\end{acknowledgments}

\clearpage

\appendix

\bibliography{apssamp}

\end{document}